\documentclass[10pt,preprint]{aastex}
\usepackage{epsf}

\newcommand\lens{J0134$-$0931}
\newcommand\vect[1]{{\bf #1}}
\newcommand\xx{\vect{x}}
\newcommand\uu{\vect{u}}
\renewcommand\aa{\vect{\alpha}}
\newcommand\ah{\hat\aa}
\newcommand\pp{\vect{p}}

\newcommand\vv{\vect{v}}
\newcommand\MM{{\rm M}}
\newcommand\mod{{\rm mod}}
\newcommand\obs{{\rm obs}}
\newcommand\src{{\rm src}}
\newcommand\kms{\mbox{km~s$^{-1}$}}
\newcommand\reffig[1]{Figure~\ref{fig:#1}}
\newcommand\reffigs[2]{Figures~\ref{fig:#1} and \ref{fig:#2}}
\newcommand\refeq[1]{eq.~(\ref{eq:#1})}

\newcommand\cp{${\rm C}'$}
\newcommand\ep{${\rm E}'$}

\begin{document}

\title{The Quintuple Quasar: Mass Models and Interpretation}

\author{Charles R.~Keeton\altaffilmark{1,2} and
  Joshua N.~Winn\altaffilmark{3,4}}
\altaffiltext{1}{Astronomy \& Astrophysics Department, University of Chicago,
  5640 S.\ Ellis Ave., Chicago, IL 60637}
\altaffiltext{2}{Hubble Fellow}
\altaffiltext{3}{Harvard-Smithsonian Center for Astrophysics, 60 Garden St.,
  Cambridge, MA 02138}
\altaffiltext{4}{NSF Astronomy \& Astrophysics Postdoctoral Fellow}

\begin{abstract}
The strange morphology of the six-component gravitational lens
PMN~\lens\ has resisted explanation. We present the first
successful quantitative models for the system, based on the idea
that there are two lens galaxies and two components of the
background source. One source is quintuply imaged and corresponds
to the five brightest observed radio components. The other source
is triply imaged and corresponds to the sixth component, along with
two others too faint to have been detected. The models reproduce
the observed image positions and fluxes, and make falsifiable
predictions about other properties of the system. Some of these
predictions have been confirmed by high-resolution radio and
optical observations, as described in the companion paper by Winn
et al.\ (2003). Although we cannot determine the lens model
uniquely with current data, we predict that the lens galaxies are
spiral galaxies with roughly equal velocity dispersions
$\sigma \sim 120\,\kms$ and a projected separation of only 0\farcs4
(2~$h^{-1}$~kpc at $z_l=0.76$). This system is the first known lens
with five images of a single quasar, and the second with more than
four images.
\end{abstract}


\section{Introduction}

The radio source PMN~\lens\ presents a formidable challenge to the
gravitational lensing theorist. Instead of having two or four
images like nearly every other lensed quasar, the system has at
least six components at radio wavelengths (Winn et al.\ 2002a;
hereafter W02). The five brightest components (named A--E in order
of radio flux density) have identical continuum radio spectra, but
C and E were missing from a high-resolution 1.7~GHz map (W02) and
also from optical and near-infrared images (Gregg et al.\ 2002;
Hall et al.\ 2002; hereafter G02 and H02). Adding to the mystery,
the object is one of the reddest quasars known ($B-K \ge 11$), but
the location of the dust presumed to cause the reddening is unknown
(G02; H02).

Confusion has reigned over such basic issues as the number of
lensed images in the system. Four completely different lensing
scenarios have been proposed. First, W02 pointed out that four of
the components (A, B, D, and F) might be explained as a 2-image
lens of a 2-component source, with C and E assumed to be foreground
objects, but one would need to invoke coincidence to explain why C
and E have the same spectral slope as A, B, and D. Second, W02
suggested that all six components could be images of a single
source if there is more than one lens galaxy, but they did not
present any quantitative models because the parameter space of
multiple-galaxy models is large and not easily explored. Third, G02
imagined A, B, D, and E to constitute a standard 4-image lens, but
admitted that such a scenario does not explain component C.
Finally, H02 suggested that A--E are five images of a single source
and F represents a different source. Although they did not present
a quantitative model, H02 pointed out two features that would be
expected of such a model: there must be at least two lens galaxies; and
there would presumably be at least one additional image of the
source corresponding to component F, which should be detectable
with more sensitive radio observations.

Here we present a comprehensive lensing analysis of \lens,
concluding that the quintuple-imaging scenario is correct, and
backing it up in quantitative detail. To do so, we originally used
modeling considerations. We exhaustively searched the parameter
space of two-galaxy lens models, finding numerous successful
5-image lens models but no good 6-image models. While this work was
in progress, Winn et al.\ (2003; hereafter W03) discovered that the
spectral index of component F is significantly different from that of the
other components. This removed any motivation to consider 6-image
models further. Furthermore, the multi-frequency maps of W03 showed
that the lower surface brightness of C and E, which was the main
objection to the H02 scenario, is due to scatter-broadening rather
than being intrinsic. We therefore present only our results for the
quintuple-imaging scenario.

Our approach was to search for mass models that could reproduce the
most obvious and most easily quantified properties of the
system---the number of radio components, their relative positions,
and to some extent their relative fluxes (\S2)---using two lens
galaxies at the same redshift with unknown positions, masses, and
shapes (\S3). Our success (\S4.1) obviated the need to consider
more complicated (and {\it a priori\/} unlikely) models involving
additional deflectors or deflectors at more than one redshift.

More interestingly, the two-galaxy models make strong predictions
about aspects of the system that are observable but were not used as
constraints. In particular, they specify the positions of the two
galaxies (\S4.2); the time delays between the images (\S4.3); the
resolved shapes of the images (\S4.4); the existence of radio arcs
(\S4.5); and the existence of two ``counter-images'' of component F
(\S4.6). New radio and optical observations have been able to test and
confirm some of these predictions, as described in the companion paper
by W03. Overall, this analysis provides a lensing context that will be
valuable for further interpretation of this complex and fascinating
system (\S5).

Throughout this paper we neglect any faint ``core'' images that
would appear near the center of the lens galaxies; see Keeton
(2003) for a general discussion of core images, and Rusin et al.\
(2001) for a discussion of the various combinations of bright and
faint images possible when there are multiple galaxies. Where
necessary we adopt a cosmology with $\Omega_M=0.3$ and
$\Omega_\Lambda=0.7$, but changing the cosmology would have little
effect on our results.

\section{Constraints on lens models}

The observed properties of \lens\ are summarized by W03. For
modeling purposes, we adopted the relative image positions and flux
ratios for components A--E from multi-frequency VLA and MERLIN maps
by W02. The astrometric uncertainties from these data are
1--2~milli-arc~seconds (mas). The VLBA maps have higher astrometric
precision, but because small-scale structure in the lens potential
can perturb the image positions by $\sim$1~mas (Mao \& Schneider 1998;
Metcalf \& Madau 2001), sub-mas precision is not desirable in an
analysis focusing on the kiloparsec-scale properties of the lens
model. Because component F has only been detected in VLBA maps, we
used the VLBA position of that component relative to D, and adopted
a 2~mas uncertainty in each coordinate.

The optical flux ratios are best avoided as model constraints
because of the contaminating influence of reddening by dust (G02;
H02; W03). Instead we used the radio flux density ratios measured
by W02. Although the statistical uncertainties in these ratios are
2--4\%, we adopted a larger uncertainty of 10\% in order to account
for possible systematic effects. One particularly important
systematic effect is small-scale structure in the lens potential;
many studies have shown that mass clumps of size
$10^{4}$--$10^{8}\,M_\odot$ can perturb radio flux density ratios
by $\sim$10\% or more (e.g., Mao \& Schneider 1998; Metcalf \&
Madau 2001; Chiba 2002; Dalal \& Kochanek 2002).

\section{Methods}

Single-galaxy models can produce lenses with more than four bright
images, but only in configurations with an even number of images
lying in a narrow annulus around the lens galaxy (Keeton et al.\
2000a; Evans \& Witt 2001), which is not the case for \lens. The
only way to produce more than four bright images in a different
configuration is with more than one galaxy (Rusin et al.\ 2001).
Our goal was to see whether two galaxies were sufficient to explain
the configuration of \lens, and if so, how well the galaxy
properties could be constrained by the data.

We modeled the galaxies as singular isothermal ellipsoids (e.g.,
Kassiola \& Kovner 1993; Kormann, Schneider \& Bartelmann 1994;
Keeton \& Kochanek 1998). This is a widely-used model because it is
consistent with the observed properties of other individual lenses,
lens statistics, and the dynamics and X-ray properties of
elliptical galaxies (e.g., Fabbiano 1989; Kochanek 1993; Maoz \&
Rix 1993; Kochanek 1996; Rix et al.\ 1997; Treu \& Koopmans 2002).
This assumption does not cause too much loss of generality; using a
different radial mass profile would mainly rescale the predicted
time delays (e.g., Kochanek 2002). In the absence of other
information, we assumed that both galaxies lie at the absorption
line redshift $z_l=0.76$ measured by H02. We also included a
possible tidal shear due to additional objects near the galaxies or
along the line of sight, because such shears seem to be common
(e.g., Keeton, Kochanek \& Seljak 1997; Holder \& Schechter 2002).

The models had 15 free parameters: the positions, ellipticities,
orientations, and masses of the two galaxies, the amplitude and
orientation of the tidal shear, and the position and flux of S$_1$,
the background source corresponding to A--E. The data provide 15
constraints: a position and flux for each of five images. We did
not include component F (or its source component, S$_2$) as a
constraint because it has no securely detected counter-images, but
instead used it as an {\it a posteriori\/} test of the models (see
\S4.6). Thus, the models formally have zero degrees of freedom.

We used the lens modeling algorithm and software written by Keeton
(2001b). For a given lens model, the software solves the lens
equation to find the predicted image positions $\xx_{i,\mod}$ and
fluxes $f_{i,\mod}$, and compares them to the observed image
positions $\xx_{i,\obs}$ and fluxes $f_{i,\obs}$ using the goodness
of fit statistic
\begin{equation}
  \chi^2 = \sum_{i=1}^{N_{\rm images}} \left[
    \frac{|\xx_{i,\mod}-\xx_{i,\obs}|^2}{\sigma_{i,x}^2} +
    \frac{(f_{i,\mod}-f_{i,\obs})^2}{\sigma_{i,f}^2} \right] ,
    \label{eq:chiimg}
\end{equation}
where the index $i$ runs over all the images, and $\sigma_{i,x}$
and $\sigma_{i,f}$ are the uncertainties in the positions and
fluxes, respectively. Models that do not produce the correct number
of images are penalized by setting $\chi^2$ to an extremely large
value. The software varies the parameters of the model, using a
downhill simplex optimization routine (e.g., Press et al.\ 1992) to
minimize $\chi^2$ and find the best fit to the data.

For this particular problem, we used two special techniques to control
the parameter search. First, to ensure a fair sampling of the large
and complex parameter space, we ran the optimization many times
starting from random points. We drew initial galaxy positions from a
box enclosing the radio components, initial ellipticities from the
range $0 \le e \le 0.6$, and random initial orientations. Second, to
speed up the optimization, we noted that it is possible to set up a
system of linear equations for the optimal values for four of the
parameters (see the Appendix).  For any set of the 11 non-linear
parameters one can simply solve the system of equations to find the
best values for the four linear parameters.  As a result, the
parameter space that needs to be searched by the optimization routine
has only 11 dimensions rather than 15.  This technique has been used
(in slightly different forms) by Kochanek (1991a, 1991b), Bernstein \&
Fischer (1999), and Keeton et al.\ (2000b).  One possible problem is
that the solutions of the linear equations actually optimize a
quantity $\chi^2_{\src}$ that is usually a good approximation to the
$\chi^2$ defined in \refeq{chiimg} but may have slightly different
minima.  Out of concern about this effect, we considered two
approaches.  In ``assisted'' models we used the linear parameter
technique throughout the modeling process, whereas in ``direct''
models we used it when selecting the random starting points but not
when optimizing the parameters.  As discussed below, the two sets of
models had similar properties, suggesting that the linear parameter
technique did not significantly bias our results.

\section{Results}

\subsection{Quality of the fits}

\begin{figure}[t]
\centerline{\epsfxsize=10.0cm \epsfbox{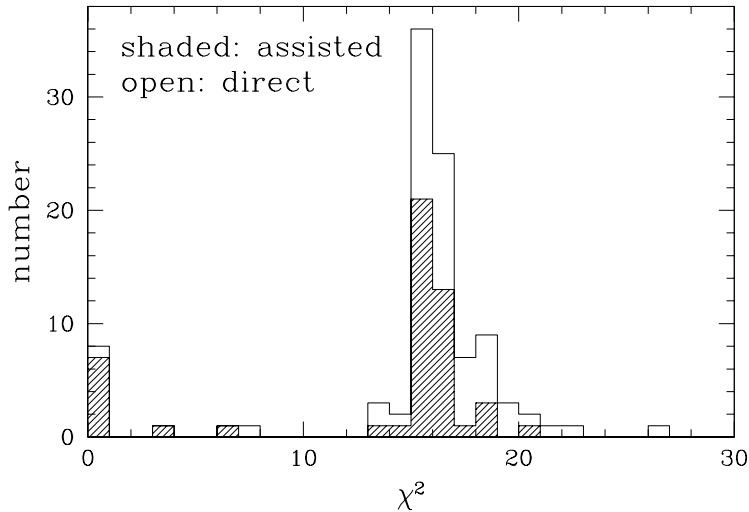}}
\caption{
Histogram of $\chi^2$ values for the models. The different
histogram indicate different types of models. There are 50 assisted
models and 50 direct models with $\chi^2<25$.
}\label{fig:chihist}
\end{figure}

We repeated our randomization and optimization modeling procedure
many times to obtain a suite of lens models that sample the local
minima in the $\chi^2$ surface. \reffig{chihist} shows histograms
of $\chi^2$ values for 50 assisted and 51 direct models with
$\chi^2<30$. In this sample, there are four assisted models that give
a perfect fit ($\chi^2=0$), and another four models (three assisted,
one direct) with $\chi^2<1$. The existence of models that fit the
data exactly is not mathematically surprising, because the models
have zero degrees of freedom. It is nevertheless interesting to
find physically plausible two-galaxy models that can account for
the configuration of this unusual lens. Our first significant
result is the mere existence of quantitatively successful models of
\lens.

There are a few models with $1<\chi^2<8$, and a large collection of
models (41 assisted, 48 direct) centered at $\chi^2 \sim 16$.
Interestingly, in the latter models most of the value of $\chi^2$
comes from the flux ratios (see Table~1). In fact, the dominant
contribution is from the flux of component A, which is commonly
predicted to be $\sim$40\% fainter than observed. This is an
important point, because recent theoretical work has indicated that
in many lenses one of the image fluxes is perturbed by mass
substructure in the lens (e.g., Mao \& Schneider 1998; Metcalf \&
Madau 2001; Chiba 2002; Dalal \& Kochanek 2002). According to such
analyses, even smooth models that correctly describe the overall
lens potential can fail to reproduce one of the flux ratios. The
smooth models will often tend to underestimate the flux of the
brightest image, which is exactly the situation for the
$\chi^2 \sim 16$ models.

Let us stress, though, that we do not claim that substructure is
required to explain the radio flux ratios in this system. That
claim is disproved by the existence of smooth models that fit the
radio data exactly. Rather, we simply argue that models with
$\chi^2 \sim 16$ dominated by a $\sim$40\% discrepancy in the flux
of A should not be ruled out. We therefore consider all models
with $\chi^2 < 25$ to be viable models. (This limit represents
a natural break in the $\chi^2$ histograms; see \reffig{chihist}.)
We study a sample of acceptable models containing 50 assisted models
and 50 direct models.

Note that there is little difference between the $\chi^2$
histograms for assisted and direct models. The assisted models appear
to be somewhat more efficient at finding models with $\chi^2 < 10$,
presumably because the linear parameter technique helps the
downhill simplex method find models that lie at the bottom of
narrow valleys in the $\chi^2$ surface (see \S3). But there is no
significant and systematic difference in the values of $\chi^2$ or
the best-fit parameter values between the two model types, which is
reassuring evidence that the linear parameter technique did not
bias our results.

\subsection{Galaxy properties}

\begin{figure}
\centerline{\epsfxsize=16.0cm \epsfbox{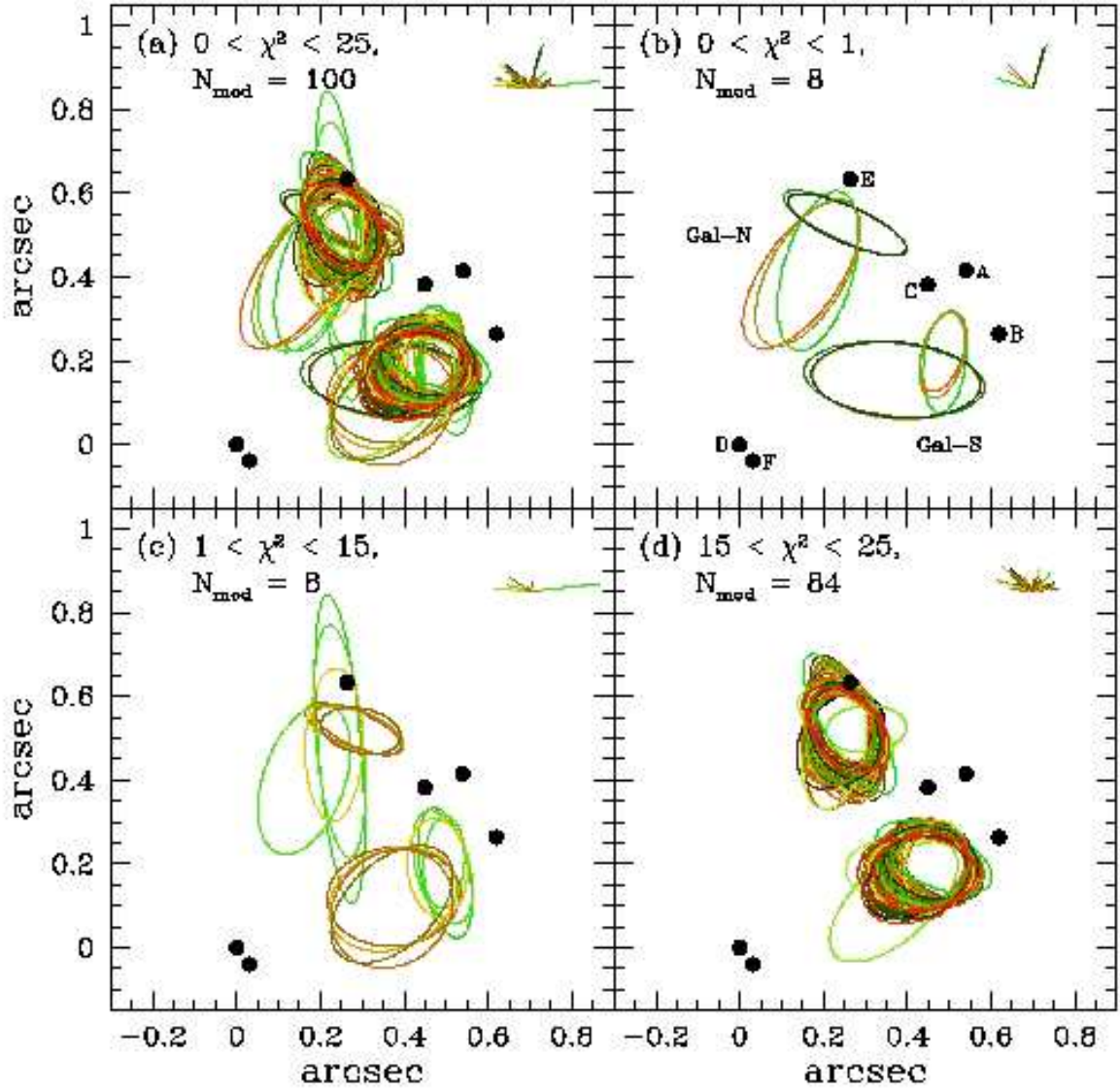}}
\caption{
Isodensity contours ($\kappa=1$) of the two lens galaxies, drawn
for various lens models. The vectors in the upper right corner
indicate the amplitude and direction of the external shear. The
points represent the observed radio components. (a) All 100
acceptable models. (b)--(d) Models in different $\chi^2$ ranges as
indicated in the key, where $N_{\mod}$ is the number of models
shown.
}\label{fig:ell}
\end{figure}

Table~1 gives typical values of the model parameters for ten
representative models, chosen to capture the range of properties seen
in the full sample of 100 models. To give a visual impression of the
range of positions, relative masses, ellipticities, and orientations
for the lens galaxies, in \reffig{ell}a we have overplotted the
$\kappa = \Sigma/\Sigma_{\rm crit} = 1$ contours for the two galaxies
in all 100 models.

The first interesting result is that the models all basically agree
on the positions of the two galaxies. The northern galaxy
(henceforth labeled Gal-N) lies $\sim$0\farcs2 south of component
C, while the southern galaxy (Gal-S) lies $\sim$0\farcs15 south of
component E. The fact that the galaxy positions vary little from
model to model indicates that they are robust predictions of
two-galaxy models.

More details become apparent when the models are grouped by their
$\chi^2$ values, as in \reffig{ell}b--d. The best models
($\chi^2<1$) fall into two different families: either the two
galaxies are both oriented nearly east--west, or they are both
oriented nearly north--south. In the range $1<\chi^2<15$, the range
of allowed ellipticities and orientations increases, and for some
models Gal-N is allowed to be highly flattened in the north-south
direction.

The models with $15<\chi^2<25$ represent the majority (84\%) of
allowed models, and form a remarkably homogeneous family that is
qualitatively different from the other models. In all but one of
these models, the galaxies are comparatively round and are located
almost directly south of of components C and E. (Recall that these
are the models that are acceptable under the hypothesis that mass
substructure affects the radio flux density of A.)

\begin{figure}
\centerline{\epsfxsize=16.0cm \epsfbox{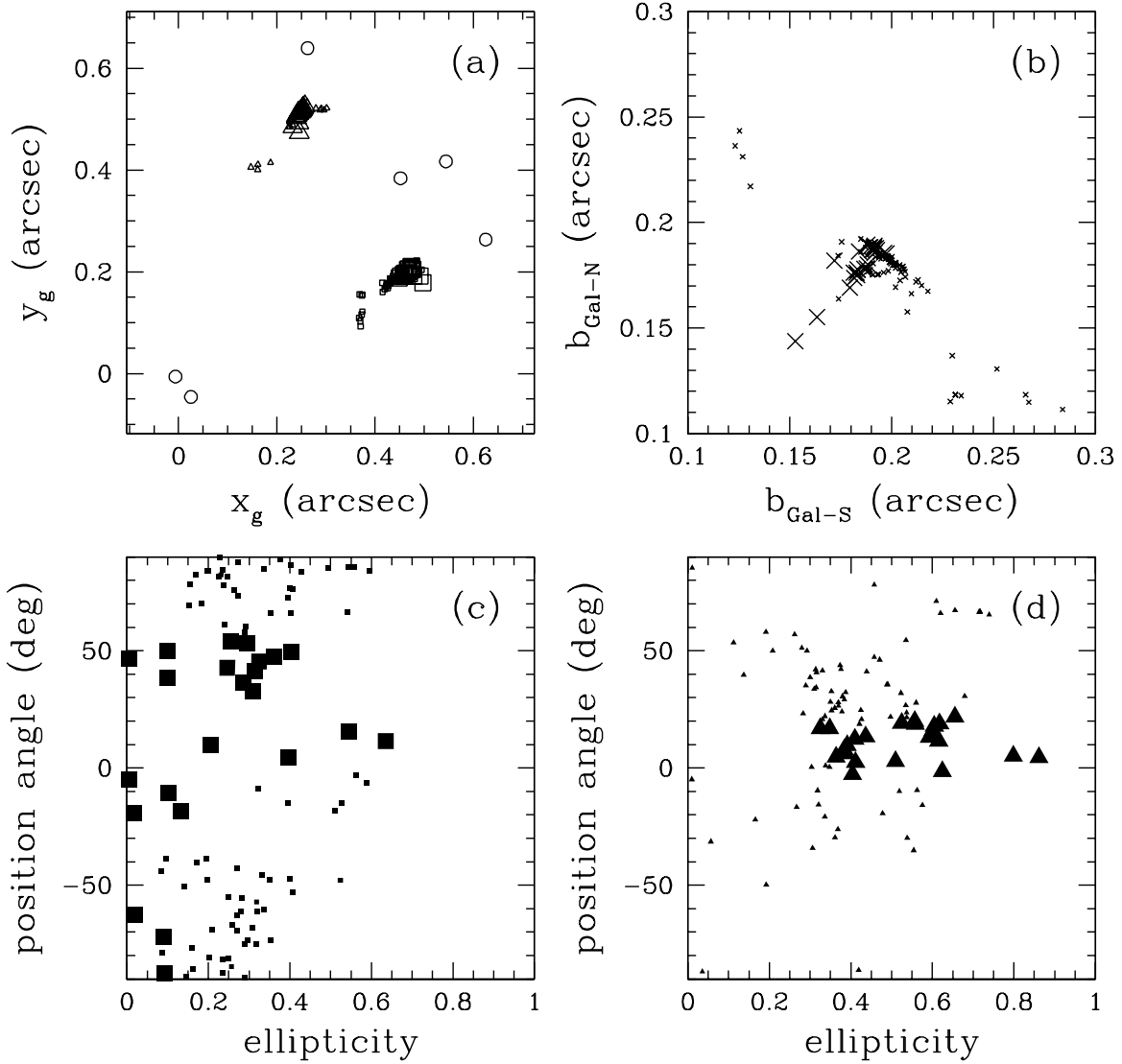}}
\caption{
Scatter plots of parameter values for acceptable models.
(a) Galaxy positions; Gal-S is shown as squares and Gal-N as
triangles. The circles indicate the positions of components A--E.
(b) The mass parameters $b_{\rm Gal-N}$ and $b_{\rm Gal-S}$.
(c) Ellipticity and position angle for Gal-S.
(d) Ellipticity and position angle for Gal-N.
In all cases, large (small) points correspond to models that are
consistent (inconsistent) with the observed 15~GHz shapes of the
images at 95\% confidence (see \S4.4).
}\label{fig:scatt}
\end{figure}

A different way to display the model results is with scatter plots
of the parameter values. \reffig{scatt}a provides this view of the
galaxy positions. As before, it is evident that the models
basically agree on the galaxy positions. There is some freedom for
the galaxies to move through a $\sim$0\farcs1 range along lines
running northwest to southeast, but as they move they must maintain
a fixed separation of $0\farcs385\pm0\farcs008$ (2.0~$h^{-1}$~kpc
at $z_l=0.76$).

\reffig{scatt}b shows the allowed values of the galaxy mass
parameters $b_{\rm Gal-N}$ and $b_{\rm Gal-S}$ (defined in
eq.~\ref{eq:bSIS} in the Appendix). In most models the two galaxies
have comparable masses, although there is some freedom to make one
or the other galaxy more massive provided that the total mass
within the region bounded by the images remains fixed. The mass
parameter $b$ can be translated into a velocity dispersion via the
relation $\sigma \approx 117\,(b/0\farcs2)^{1/2}\,\kms$, assuming a
lens redshift $z_l=0.76$ and neglecting an ellipticity-dependent
factor of order unity.

Finally, Figures~\ref{fig:scatt}c and \ref{fig:scatt}d show that there
is great diversity in the allowed ellipticities and orientations of
the two galaxies.  The models fill a broad region in these slices of
parameter space.  We therefore cannot determine these aspects of the
lens models from the image configuration alone, although some
discrimination may be possible from observations of the intrinsic,
resolved shapes of the individual images (see \S4.4).

When G02 deconvolved a ground-based $K'$-band image, they found two
faint components in addition to A, B, and D. They suggested that,
if these components were real and not deconvolution artifacts, they
might correspond to component E and a lens galaxy. Because the
positions of these peaks agree well with the two predicted galaxy
positions, we suggest instead that G02 detected both of the lens
galaxies.

Furthermore, H02 identified possible flux from a lens galaxy or
galaxies in Hubble Space Telescope (HST) images but were unable to
draw any detailed conclusions. In a more careful analysis of the
HST images that was conducted in tandem with this work, W03 found
clear evidence for two flux peaks. The peaks agree roughly with the
positions predicted by the models, although imperfect PSF
subtraction makes it difficult to measure their positions
accurately.

Thus, the $K'$-band and optical HST images appear to confirm
the first important prediction of the models, the presence of two
galaxies near components C and E. It is interesting to note that
the southern peak observed by W03 in the HST images appears to be
elongated in the east--west direction. In principle, one might use
this information to rule out the broad class of models that predict
Gal-S to be round or elongated north--south. However, we hesitate
to rule out models based on the HST images, not only because the
residual peaks are faint and poorly characterized, but also because
the mass distribution need not follow the light distribution.
Differences between the mass and light distributions would not be
surprising of the two galaxies are interacting, or if they are
spiral galaxies (as we suspect) with prominent spiral arms.

\clearpage

\subsection{Components A--E}

\begin{figure}
\centerline{
  \epsfxsize=7.5cm \epsfbox{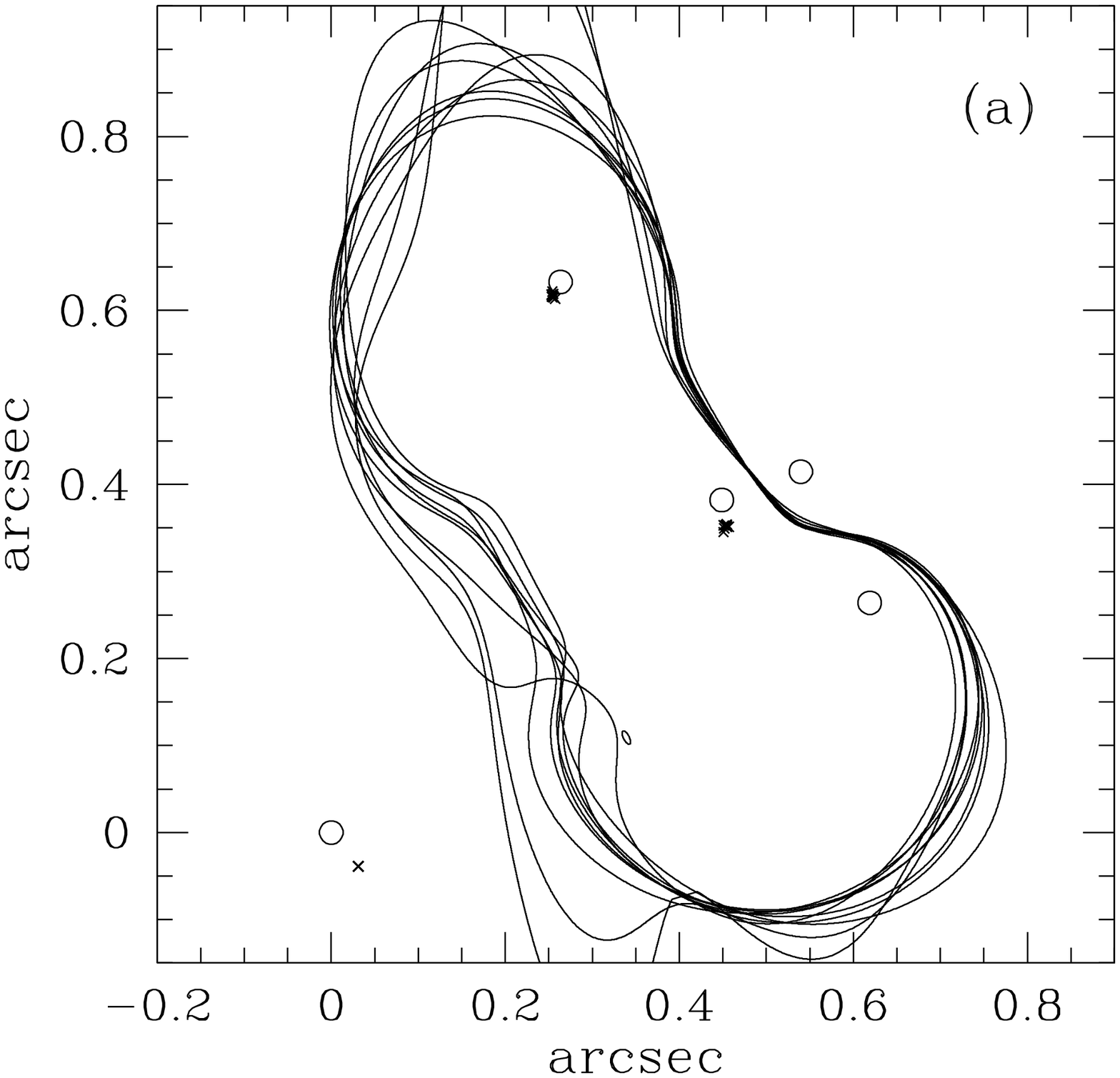}
  \epsfxsize=7.5cm \epsfbox{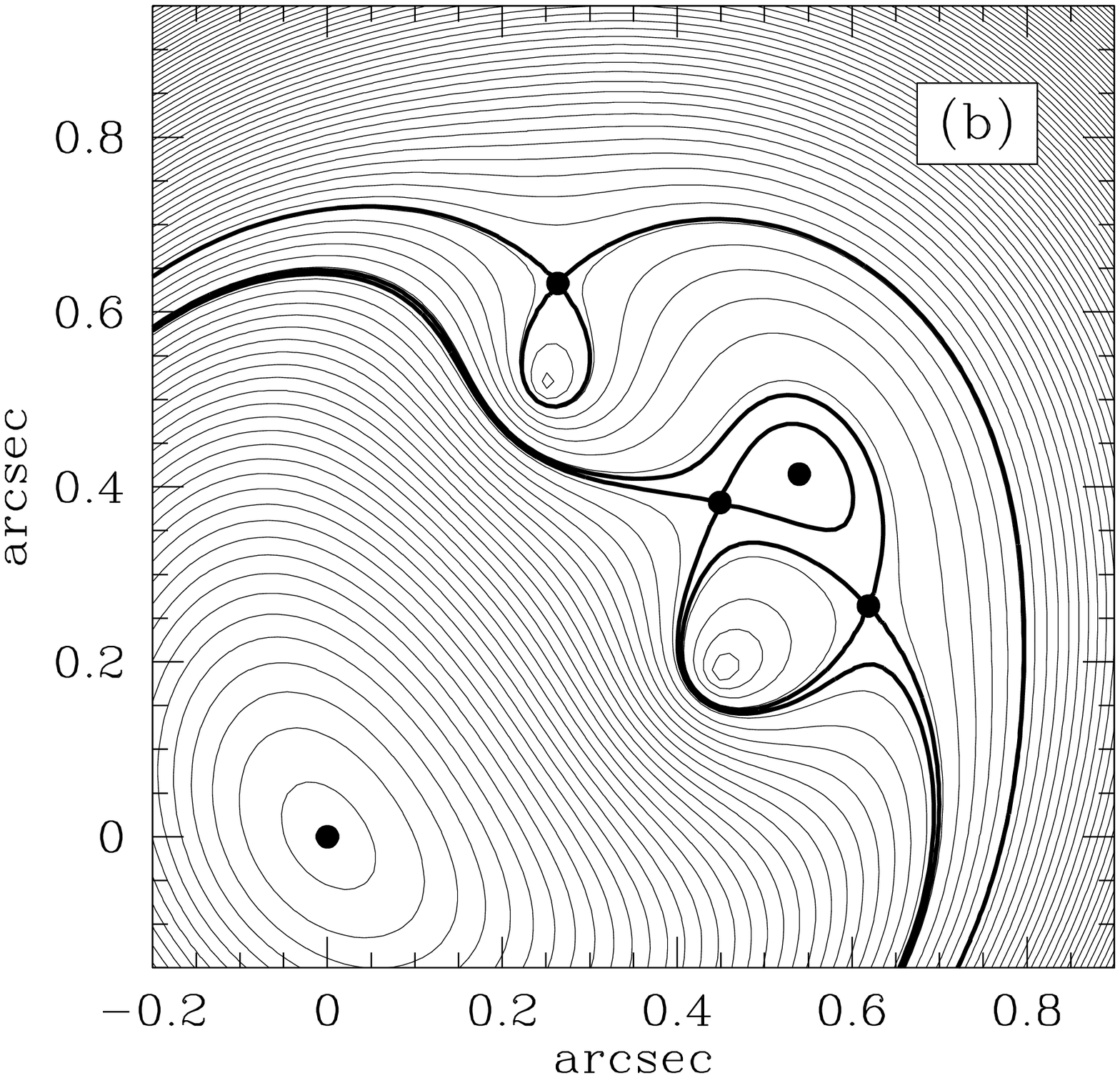}
}
\centerline{
  \epsfxsize=7.5cm \epsfbox{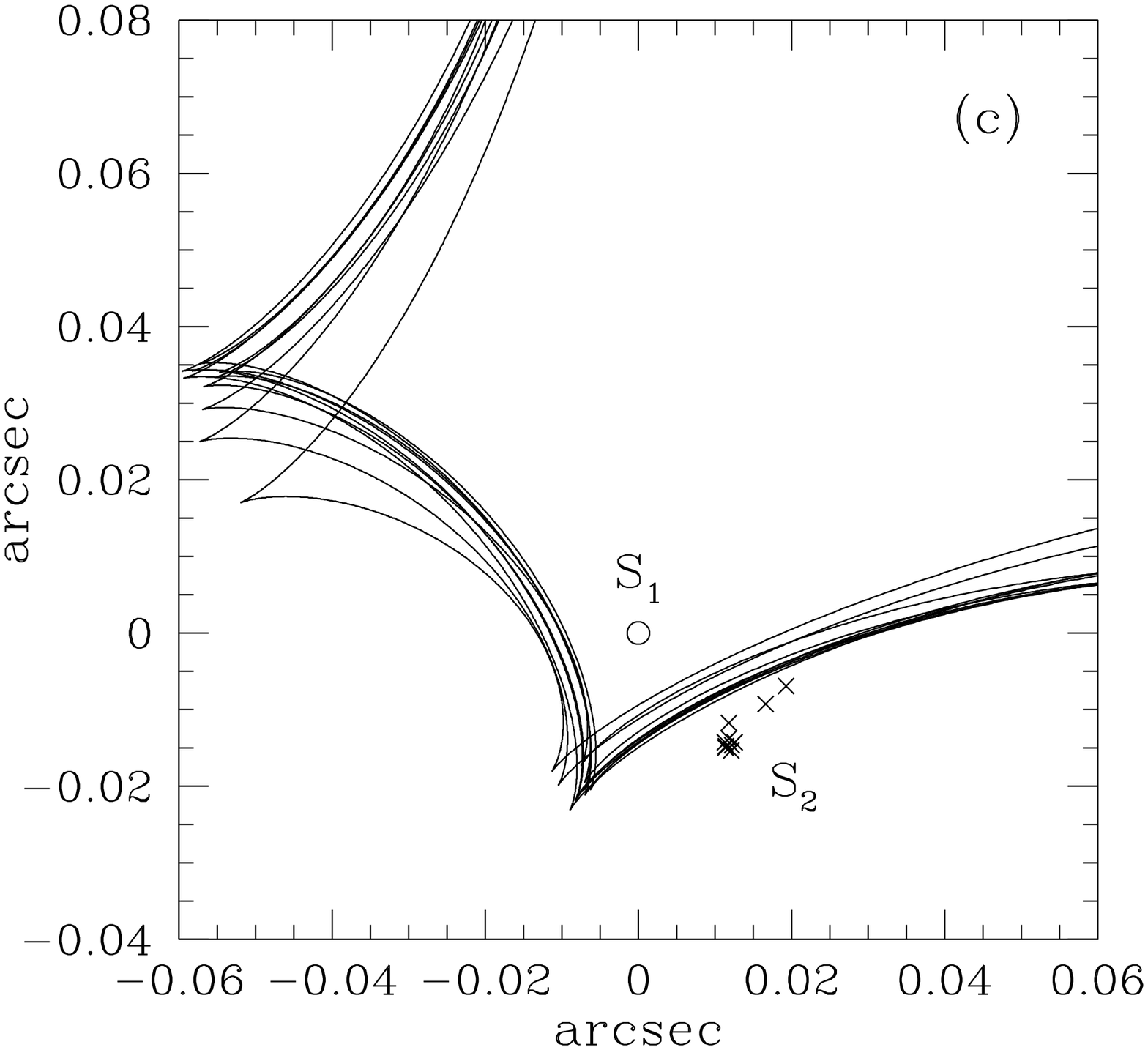}
  \epsfxsize=7.5cm \epsfbox{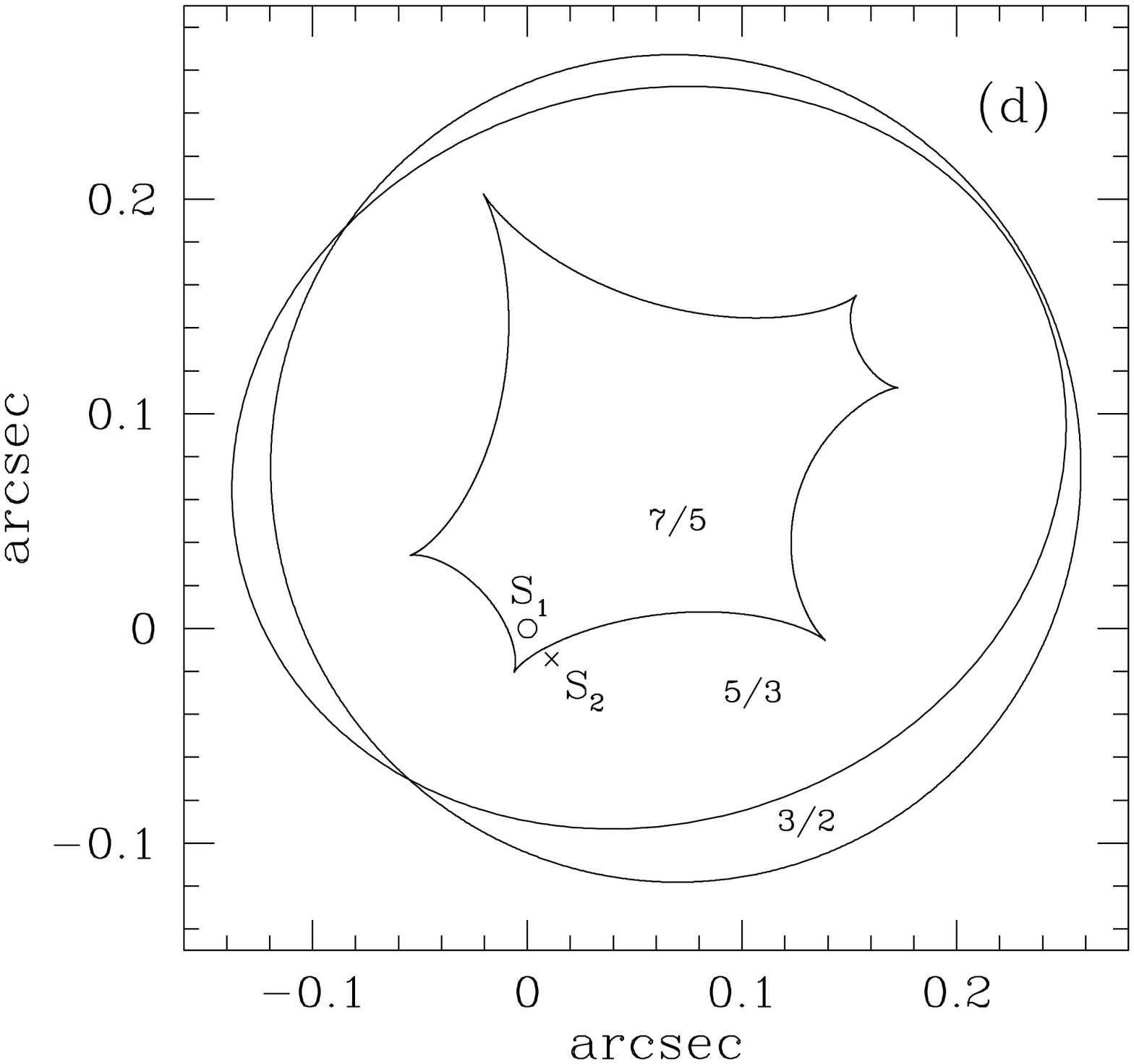}
}
\caption{
(a) Critical curves for the ten models in Table~1. The circles show
the positions of components A--E, while the crosses indicate the
positions of component F and its predicted counter-images.
(b) Time delay surface for a typical model.  The light contours
are spaced by 0.25 $h^{-1}$ days, while the heavy contours are
drawn to pass through the saddle points (images B, C, and E).
(c) Caustics for the ten models in Table~1, in coordinates centered
on S$_1$ (circle). The crosses indicate the position of the source
component S$_2$.
(d) Full caustic structure for a typical model.  The numbers in the
form $N_{\rm tot}/N_{\rm bri}$ show the total number of lensed images
$N_{\rm tot}$ and the number of bright (non-core) images $N_{\rm bri}$
for sources in the different regions of the source plane.
}\label{fig:crit}
\end{figure}

The lensing properties of this system can be understood by examining
the critical curves, caustics, and time delay surface, which are
shown in \reffig{crit}. (To avoid clutter only the ten models listed
in Table~1 are shown, but the other models are qualitatively similar.)
Although there is some freedom for the critical curve to meander,
especially to the southeast near component D, the data require the
critical curve to thread between A, B, C, and E along a
well-constrained path. Components A and D both lie outside the
critical curve and at minima of the time delay surface, so they
are predicted to have positive parity; while B, C, and E all lie
inside the critical curve and at saddle points of the time delays
surface, so they are expected to have negative parity. These
results have implications for the resolved shapes of the images
(see \S4.4).

The models all predict that S$_1$, the source corresponding to
components A--E, lies only $\sim$0\farcs02 from a cusp in the
caustic (\reffig{crit}b). The models also generally agree on the
position and orientation of that cusp, and on the presence of a
second cusp nearby. There is less agreement about the shape of
the caustic farther from the source, which corresponds to the
poorly-constrained stretch of the critical curve. The fact that
the source lies so close to a cusp means that the quasar host
galaxy might be visible as a partial or complete Einstein ring
in a high-resolution near-IR image of the system (see Kochanek,
Keeton \& McLeod 2001).

One feature of \lens\ cited by W02 as circumstantial evidence
for gravitational lensing was the fact that four of the radio
components (A, B, D, and F) lie on a circle, and that the A/B
radio arc conforms closely to the same circle. However, as our
models do not involve an intrinsically circular mass distribution,
we suggest that the near-perfect circularity is just a mild
coincidence. Any three points (say, A, B, and D) define a circle.
Any extended emission should be seen most prominently between
A and B because they are the brightest images. The extended
emission, along with any additional images near D, are likely to
lie in the direction of maximum stretching in the image plane,
which is roughly perpendicular to the A--D
separation.

\begin{figure}[t]
\centerline{\epsfxsize=10.0cm \epsfbox{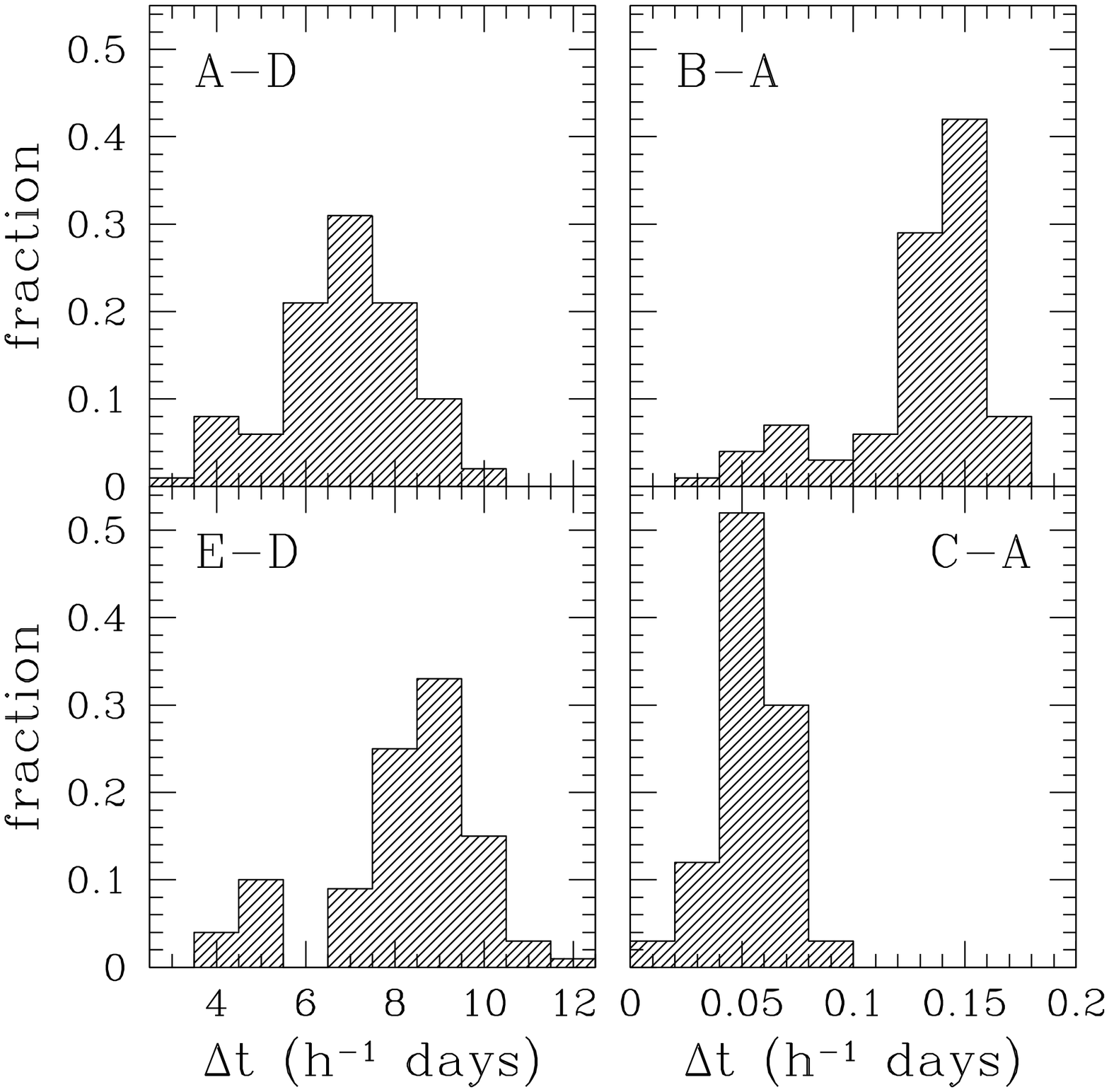}}
\caption{
Histograms of the predicted time delays between various image
pairs.
}\label{fig:tdel}
\end{figure}

The differential time delays between the images can be predicted
from the models. In all cases the temporal ordering of the images,
from shortest time delay to longest, is D--A--C--B--E. The delay
values are shown in \reffig{tdel}, using the source redshift
$z_s=2.2$ and assuming that the galaxies lie
at redshift $z_l=0.76$. The delays are short because the angular
size of the lens is small. The D--A delay is only $\sim$10 days
($\sim\!7\,h^{-1}$ days), and the A--E delay is only another
$\sim$2 days. The small separations between components A, B, and
C mean that the time delays between them are just a few hours.

Little or no variability has been measured in the total radio flux
density over a 20-year baseline (W02), which is reason for pessimism
about the prospect of measuring the time delays at radio
wavelengths. One might hope for a greater degree of variability at
optical or near-infrared wavelengths. The time delays would be
challenging to measure at these wavelengths because of the small
angular size of the system, but G02 demonstrated that it is possible
to resolve some of the components in ground-based images with
reasonable accuracy. Another possibility is to attempt to measure the
short delays between A, B, and C in a single X-ray observation; the
general feasibility of this approach has been demonstrated by Dai et
al.\ (2001), but it would depend on the X-ray flux of the quasar.
Although the complexity of \lens\ makes it unlikely ever to be useful
for Hubble constant determination, measurements of the time delays
would provide valuable new constraints on lens models.

\subsection{Resolved image shapes}

In high-resolution VLBA observations, W03 resolved the five radio
components A--E.  Although we did not use this information to derive
our lens models, we can use it as an {\it a posteriori\/} test of the
models.  Specifically, we considered the images shapes inferred from
elliptical Gaussian fits to the 15~GHz VLBA maps. We chose 15~GHz to
minimize the possible effect of scatter-broadening, and we used only
the position angles (rather than the major and minor axes) because the
true 15~GHz shapes are not simply ellipses. For modeling purposes, we
used an elliptical source, because it can easily be combined with the
lensing magnification tensor to estimate the orientation of the lensed
image (Keeton 2001a). It ought to be reasonable to compare predicted
orientations of images inferred from an elliptical source to the
measured orientations, even if the images are not strictly elliptical.

For each of the 100 lens models we varied the axis ratio and
orientation of the source component S$_1$ to find the best fit. The
best case has $\chi^2_{\rm shape} = 0.58$ for three degrees of freedom.
A total of 23 models are consistent with the data at better than 95\%
confidence ($\chi^2_{\rm shape} < 7.8$ for three degrees of freedom).
These models all have sources with an axis ratio between 1.5 and 2.0
and position angle between $145\arcdeg$ and $153\arcdeg$.

The fact that 77 of our otherwise successful models are formally
excluded by the image shapes indicates that the shapes contain
important higher-order information about the lensing potential. It
reassures us that the remaining 23 models are reasonable descriptions
of the potential. It also narrows the range of allowed lens galaxy
properties. \reffig{scatt} shows that the models that agree with the
images shapes occupy smaller regions of parameter space than the full
set of models. Specifically, the two galaxies must lie at the western
end of the allowed range of positions (and hence due south of
components C and E), and Gal-N must be moderately elongated
north--south. Interestingly, the image shapes seem to exclude models
in which Gal-S is elongated east--west, suggesting that its mass
distribution may not follow its light distribution (see \S4.2).

We caution, however, against over-interpreting the $\chi^2_{\rm
shape}$ values because fitting the complex 15~GHz image shapes with
elliptical Gaussians probably underestimates the uncertainties in the
position angles (see W03). We consider the models that produce formal
agreement to be favored, but do not consider the other models to be
ruled out definitively.

\subsection{Extended emission}

\begin{figure}[t]
\plottwo{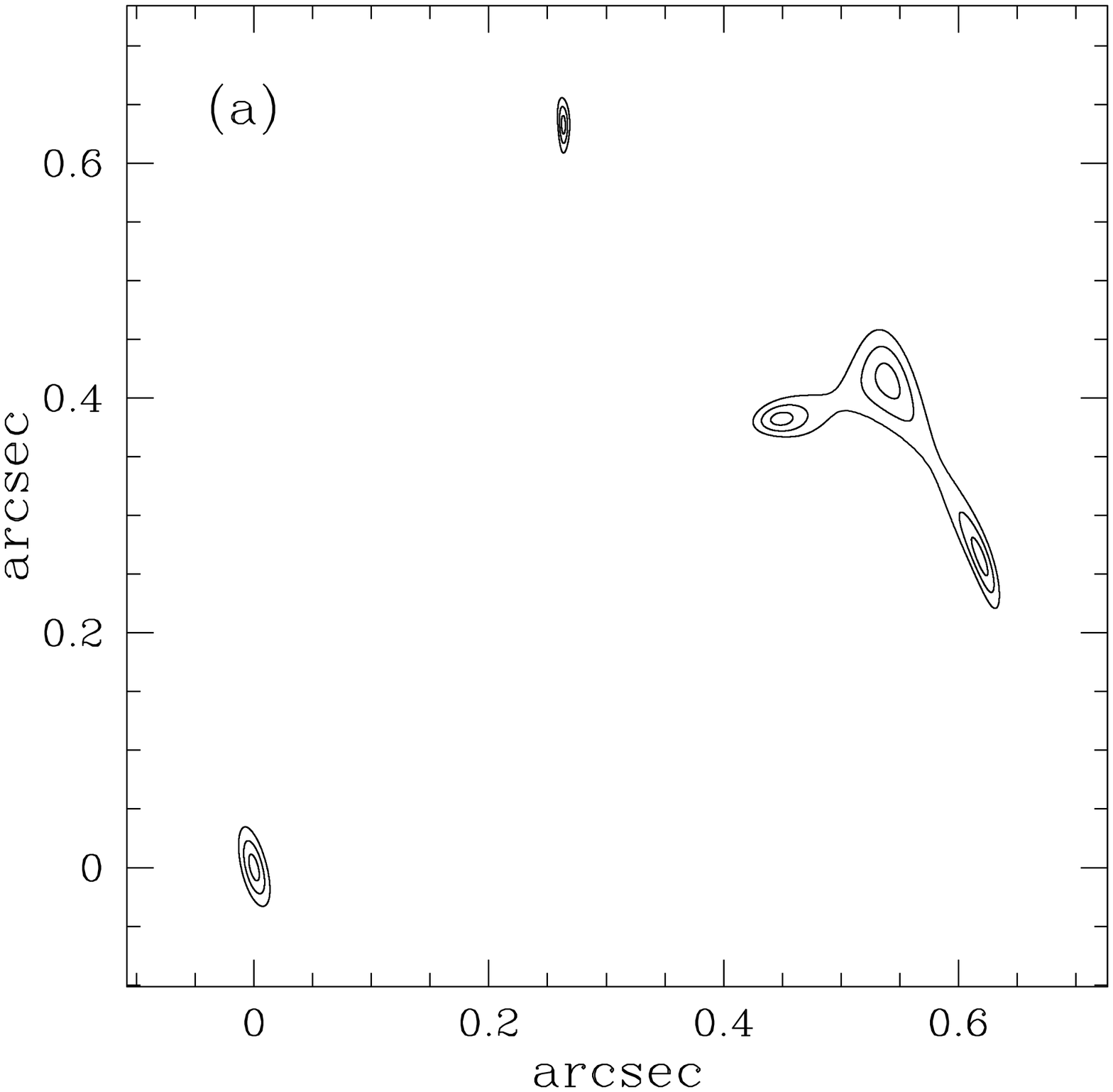}{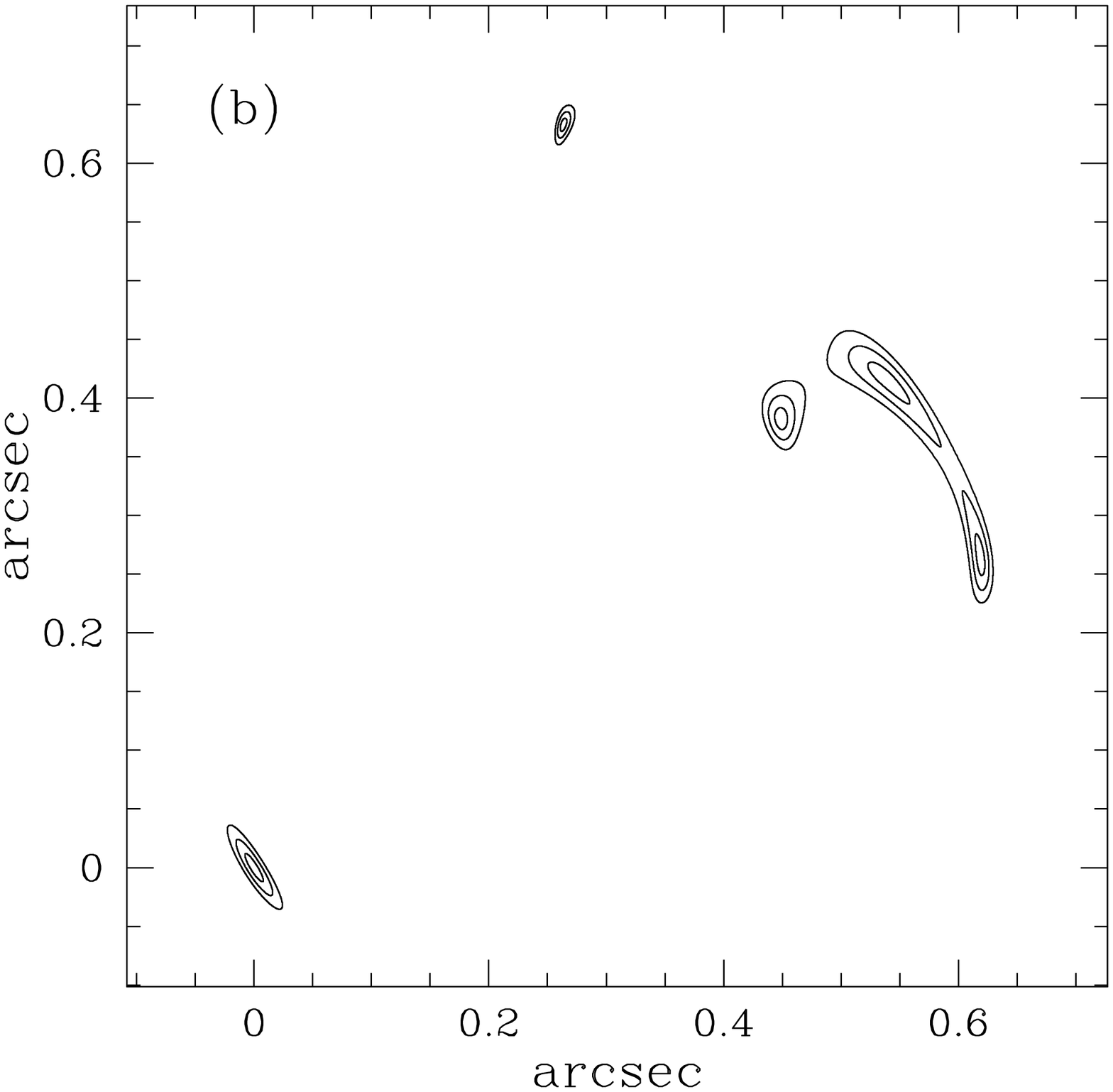}
\caption{
Images of an elliptical source centered on S$_1$ with axis ratio
1.8 and position angle $148\fdg 5$ (panel a) or $58\fdg 5$
(panel b).  The contours correspond to isophotes at 0\farcs005,
0\farcs010, and 0\farcs015 along the major axis.
}\label{fig:arc}
\end{figure}

W02 detected a curved arc of radio emission at 1.7~GHz between
components A and B.  Again, because we did not use this information
to derive our lens models, we can use it as an independent check of
the models.  At 15~GHz, the angular sizes of A--E are no greater
than a few milli-arcseconds and no arcs are seen (W03), both of
which imply an angular size of $\sim$1~mas for S$_1$.  At lower
frequencies, though, one might expect the source to be larger and
any extended emission (which typically has a steeper radio spectrum
than the core) to be more prominent.  As a first step, we took
the model that gives the best fit to the 15~GHz image shapes and
enlarged the angular size of S$_1$ to 15~mas.  The result is shown
in \reffig{arc}a.

The model predicts extended emission between components A and B,
and also between components A and C.  These two predictions appear
to be robust and common to models that reproduce the 15~GHz image
shapes, although the detailed surface brightness profiles of the
A/B and A/C extended emission depend on the angular extent,
orientation, and shape of S$_1$.  The bright A/B arc discovered by
W02 is therefore compatible with the models.  In addition, W03 found
evidence for a connection between components A and C.

However, \reffig{arc}a is not the best possible fit
to the surface brightness distributions observed by W03.  The model
predicts the A/B emission to lie nearly along a straight line, but
it is actually curved along a nearly circular arc.  In addition, the
model predicts the A/C emission to have nearly the same brightness
as the A/B emission, but it is apparently fainter (although scatter
broadening at 1.7~GHz may contribute to the difference).

Trying to optimize the model to fit the extended emission would
require a detailed description of the surface brightness distribution
plus a much more sophisticated modeling algorithm, and is beyond the
scope of this paper.  Instead, we simply note that a better fit to
the data is obtained when S$_1$ is rotated by $90\arcdeg$.  The
results for that case are shown in \reffig{arc}b.  The A/C arc is now
fainter and the A/B arc is curved, as observed.  This model has the
additional virtue that the major axis of S$_1$ points nearly in the
direction of S$_2$, which is what one would expect for a core-jet
source.  In this scenario, S$_2$ is neatly explained as a hot-spot in
the $\sim$15~mas jet emerging from S$_1$.

Taken at face value, the analyses of the 15~GHz image shapes and the
1.7~GHz extended emission suggest that the jet is bent by $\sim$90
degrees between its emergence from S$_1$ on scales of $\lesssim$1~mas
and its continuation to S$_2$ on scales of $\sim$15~mas.  This
might seem unnatural, but in fact bends of this size have been
observed within the inner few milli-arcseconds of active galactic
nuclei (e.g., Kellerman et al.\ 1998), and are usually explained as
a smaller three-dimensional bend angle viewed nearly along the jet
axis.

\subsection{Component F}

\begin{figure}[t]
\centerline{\epsfxsize=10.0cm \epsfbox{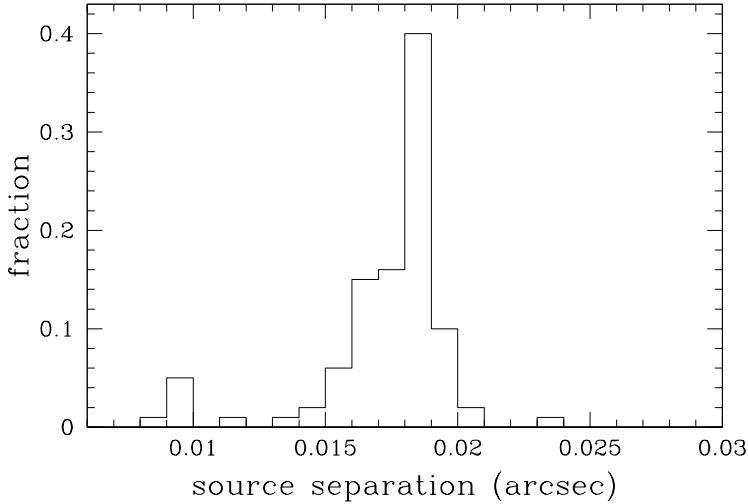}}
\caption{
Histogram of the inferred separation between the two sources
S$_1$ and S$_2$.
}\label{fig:srcsep}
\end{figure}

\begin{figure}[t]
\centerline{\epsfxsize=10.0cm \epsfbox{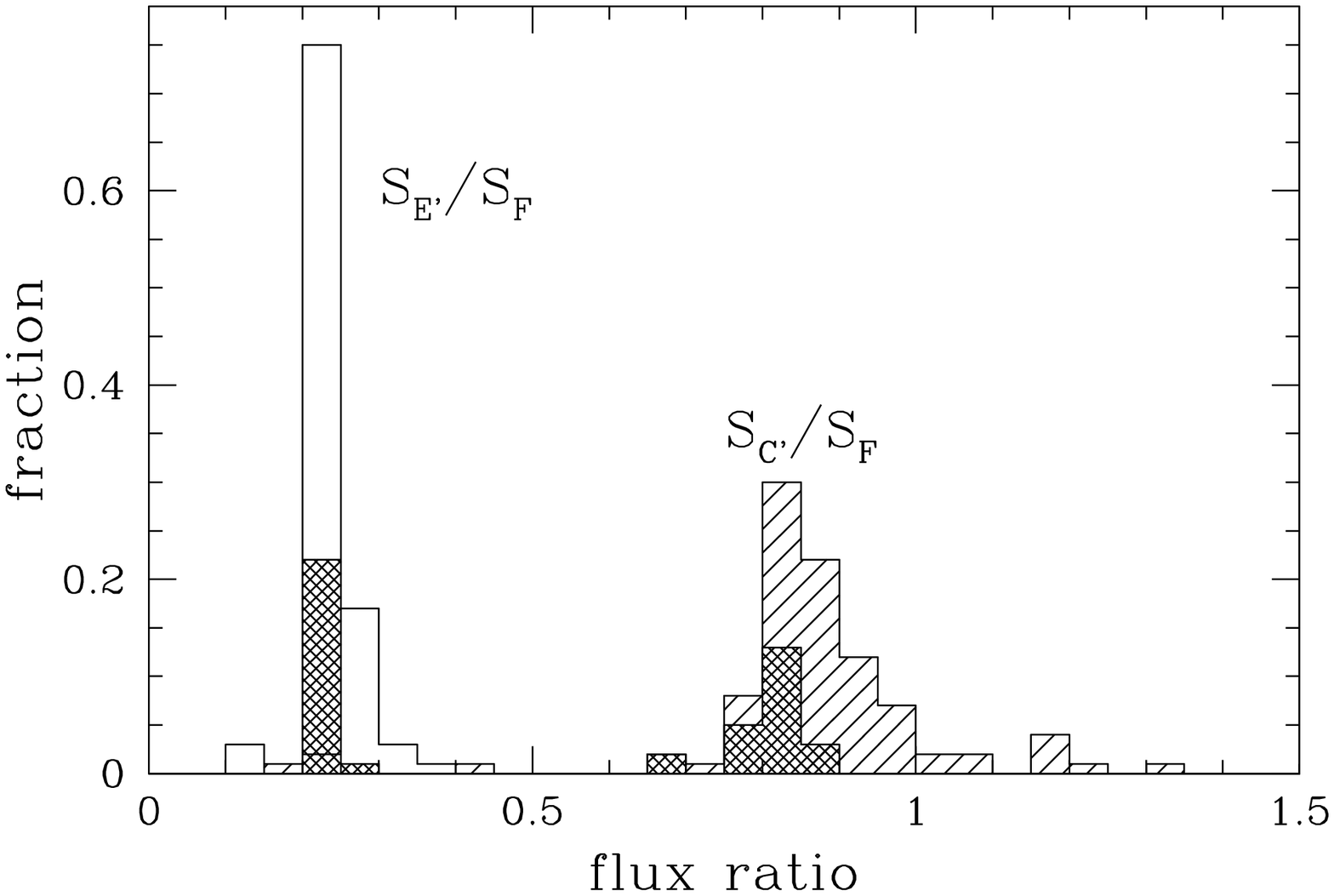}}
\caption{
Histograms of flux ratios (relative to F) for the predicted
counter-images of F. The open and light shaded histograms denote
\ep\ and \cp, respectively, for our full set of
100 models. The heavy shaded histograms are for the subset of
models that are consistent with the 15~GHz image shapes.
}\label{fig:Frat}
\end{figure}

Component F, with its steeper spectral index (W03), corresponds to
a different source component in our models that we call S$_2$.
According to the models there should be additional lensed images of
S$_2$ that have not yet been detected. To learn about these
counter-images of F, we used each of the 100
allowed models to map F into the source plane, locating S$_2$. We
then solved the lens equation for S$_2$ to find the positions and
fluxes of all images besides F. \reffigs{crit}{cimg} show the
positions of S$_2$ and the counter-images. \reffig{srcsep} is a
histogram of the separation between S$_1$ and S$_2$. Finally,
\reffig{Frat} shows the predicted fluxes of the counter-images
relative to F.

\begin{figure}[t]
\centerline{
  \epsfxsize=8.0cm \epsfbox{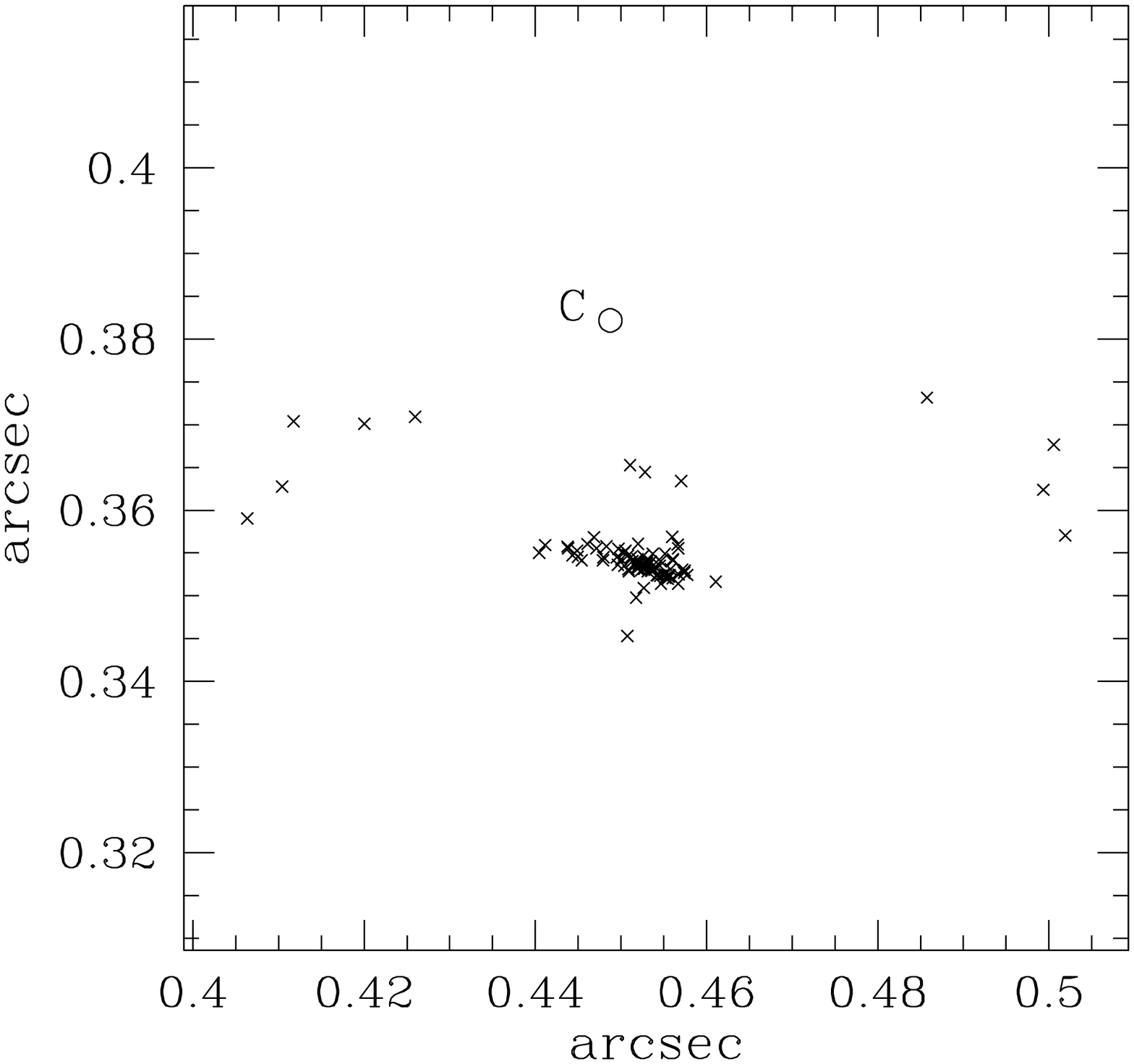}
  \epsfxsize=8.0cm \epsfbox{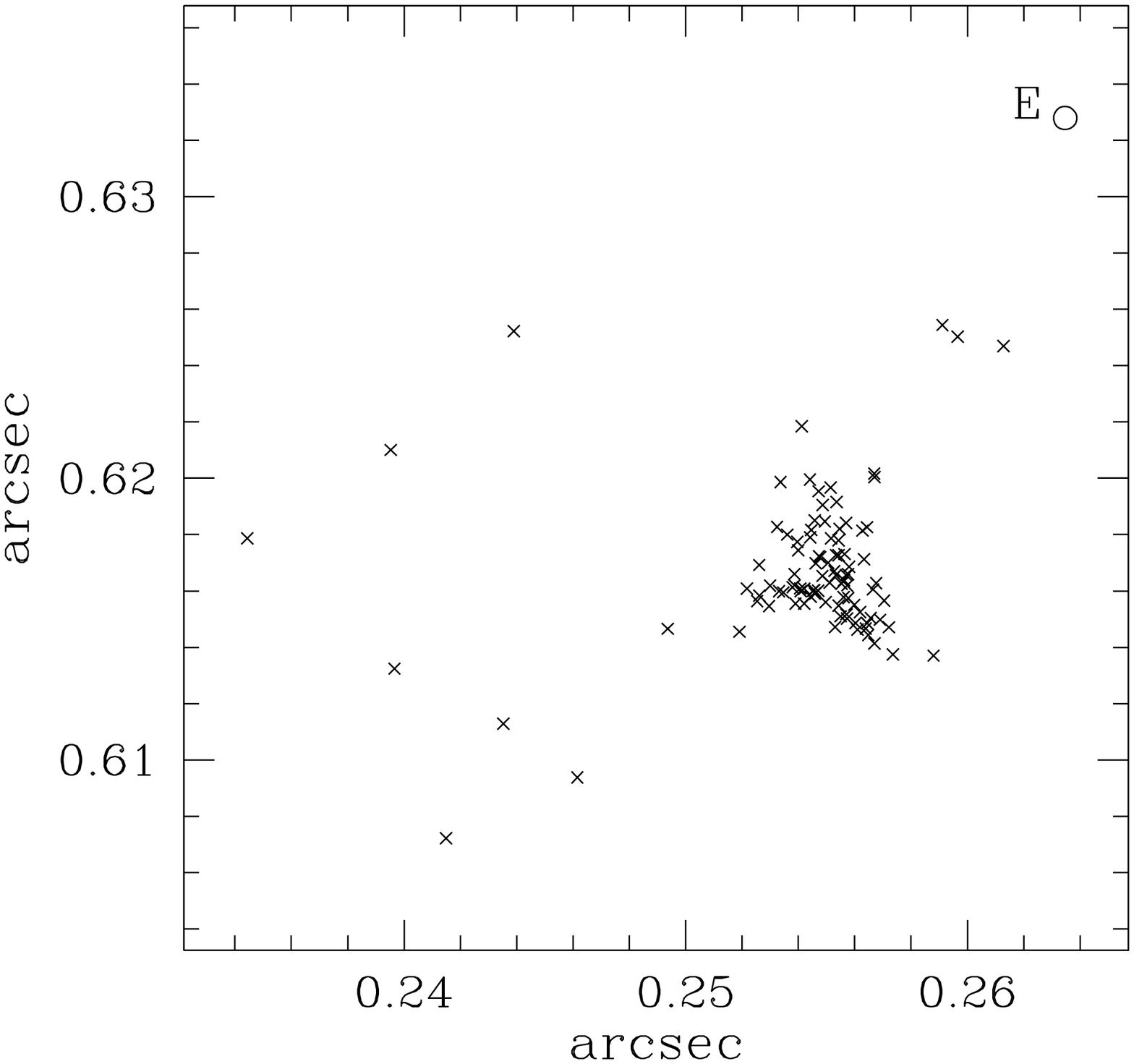}
}
\caption{
Predicted positions of the counter-images of F near component C
(left) and E (right). The circles show components C and E, while
the crosses show the predicted counter-images \cp\ and \ep\ for
the 100 acceptable lens models.
}\label{fig:cimg}
\end{figure}

All but one of the models predict that S$_2$ lies just outside the
cusp caustic and hence produces three images. Thus, F should have
two counter-images that lie close to components C and E. It is
striking that even though F lies only 0\farcs05 from D, and even
though S$_2$ is only $\sim\!0\farcs015$--$0\farcs020$ from S$_1$,
and even though no information about F or its counter-images was
used to derive the models, nearly all of the models predict that F
belongs to a family of images with a different multiplicity than
A--E. In the one exception, S$_2$ lies just inside the cusp caustic
and F has four counter-images: one near C, one near E, and a pair
straddling the critical curve between A and B.
The existence of (at least) two counter-images of F is therefore
a robust and testable prediction of the models.

We refer to putative counter-images near C and E as \cp\ and \ep,
respectively. We reserve the names G and H, which would follow from
the flux-order convention by which A--F are named, for any actual
detections of these components. This is partly because the
predicted flux ordering is not certain, although generally F is
predicted to be brighter than \cp, which in turn is brighter than
\ep. In the majority of cases, \cp\ is 80--90\% as bright as F
(especially for models that agree with the measured 15~GHz image
shapes; see \reffig{Frat}), while \ep\ is only $\sim$25\% as bright
as F.

Given that \cp, in particular, is predicted to be only modestly
fainter than F, one might expect that it could easily be detected
or ruled out. Unfortunately, circumstances make detection of the
counter-images difficult and obscure the interpretation of
non-detections. Because of the steep radio spectrum of S$_2$, one
would prefer to search for \cp\ and \ep\ at low radio frequencies
where their flux densities are large. However, at low radio
frequencies, C and E are heavily scatter-broadened, suggesting that
\cp\ and \ep\ might also be broadened out of detectability. Given
the high electron column density implied by the scatter-broadening,
it is also possible that the flux of these components is being
reduced by free-free absorption. At high radio frequencies, where
these plasma effects are smaller, the counter-images are predicted
to be intrinsically faint.

These challenges are illustrated by the analysis of new
high-resolution VLBA maps by W03. In the 1.7~GHz map, component C
is missing, because of scatter-broadening and possibly free-free
absorption. Intriguingly, however, there is a puff of resolved
emission located just south of the expected position of C, which is
exactly where \cp\ is predicted to appear. The total flux of the
extended emission is $\approx$90\% of the flux of F at that
frequency, which is also consistent with the prediction for \cp. It
is therefore possible that W03 detected \cp. But if the figure of
90\% is correct, then W03 should also have detected \cp\ at 8.4~GHz
(where it should have been much more compact due to the
$\nu^{-2}$-dependence of scatter-broadening) and they did not.

If \cp\ and \ep\ were point sources at 8.4~GHz, and if
scatter-broadening were neglected, the 5$\sigma$ upper limits of
the map would imply $S/S_{\rm F}<0.57$ for both counter-images. The
limit on \ep\ is consistent with all the models, but the limit on
\cp\ formally excludes 96\% of our otherwise acceptable models,
including all the models that agree with the 15~GHz image shapes.
However, if \cp\ and \ep\ were scatter-broadened even at 8.4~GHz,
the upper limits would be weakened. For example, under the
assumption that \cp\ and \ep\ have the same ratio of peak to total
flux density as C and E, respectively, the 8.4~GHz upper limits
would become $S_{{\rm C}'}/S_F<2.1$ and $S_{{\rm E}'}/S_F<1.1$,
which are consistent with all of our models.

Thus, it appears that the new observations by W03 were not
conclusive in this regard. This is unfortunate, because the
detection of a counter-image would be a clear validation of the
models. Probably the best strategy for future attempts to detect
the counter-images is to conduct more sensitive VLBI observations
at around 8~GHz, where the plasma effects are expected to be small,
and the intrinsic flux density of S$_2$ is still fairly large. If
the sensitivity could be improved by a factor of $\sim$3, then \cp\
should be revealed; if by a factor of $\sim$10, then \ep\ should
also be revealed. However, because the plasma effects cannot be
quantified without assumptions about the density, temperature, and
path length through the ionized material, a non-detection would be
difficult to interpret unless one worked at 15~GHz or higher, where
the observed angular sizes of C and E appear to be mainly
intrinsic.

\section{Discussion}

The observed properties of the unique gravitational lens \lens\ can be
completely explained as follows. The background radio source has two
components, S$_1$ and S$_2$, separated by $\sim$15--20~mas
(90--120~$h^{-1}$~pc at $z_s=2.2$). Source S$_1$ is the radio core with
a gigahertz-peaked radio spectrum and a quasar optical counterpart
(W02; G02).  Source S$_2$ has a steeper radio spectrum (W03) and is
a hot-spot in a jet emerging from S$_1$.  The jet is bent, turning
by $\sim\!90\arcdeg$ from its emergence from S$_1$ ($\lesssim$1~mas)
to its continuation towards S$_2$ ($\sim$15~mas).

There are two lens galaxies in the foreground producing five images of
S$_1$ (observed as radio components A--E) and three images of S$_2$
(of which only F has been securely detected, although W03 found
evidence for one additional image).  The $\sim$10~mas jet between
S$_1$ and S$_2$ is magnified into the extended emission observed at
low frequencies between A and B, and between A and C.  The inner
$\lesssim$1~mas of the jet is displayed in the resolved image shapes
at 15~GHz.

The lens galaxies have nearly equal mass ($\sigma \sim 120\,\kms$) and
are seen in projection just south of components C and E. Due to
ionized material in the northern galaxy, components C and E (and, to a
lesser degree, component A) are being scatter-broadened at radio
wavelengths (W03). Due to differential reddening by dust in both
galaxies, the optical counterparts of the images have different colors
(H02; W03). The inference of large amounts of dust and ionized
material leads us to suspect that the galaxies are spiral galaxies.

We have shown that this scenario can account quantitatively for the
observed positions and relative fluxes of the radio components. In
addition, W03 have presented evidence for the direct detection of
the lens galaxies in HST images, in agreement with an earlier
detection of two flux peaks by G02 in a $K'$-band deconvolution. A
subset of our models are also compatible with the intrinsic image
orientations measured by W03, and with the upper limits on the flux
densities of the counter-images of component F (although there are
potential ambiguities in the interpretation of both constraints).

Two puzzles remain. First, many of our models, and especially those
that are consistent with the 15~GHz image shapes, predict that the
radio flux density of component A should be $\sim$40\% lower than
observed. It is possible that a small-scale density fluctuation in
the lens model near image A is enhancing its flux density via the
substructure lensing effect (e.g., Mao \& Schneider 1998; Metcalf \&
Madau 2001).  The density fluctuation might represent a mass clump
such as a dark matter subhalo or globular cluster, or perhaps a tidal
tail if the two lens galaxies are interacting.  Although we cannot
test this hypothesis with the present analysis, it may be possible
to do so with a detailed study of the intrinsic shape of image A
(see Metcalf 2002).

Second, our models suggest that the mass in the southern galaxy
(Gal-S) cannot be highly elongated in the east--west direction and
still be consistent with the observed 15~GHz image shapes.  The puzzle
is that there are two indications that Gal-S is indeed elongated
east--west: the HST detection of Gal-S appears to be elongated
east--west (W03); and component D is highly reddened (H02; W03),
suggesting it is being covered by an easterly extension of Gal-S.
Taking all this evidence seriously requires that the light and dust do
not follow the mass. Although this might be unexpected, it does not
demand an exotic explanation.  There may simply be a spiral arm
covering D.  Or perhaps Gal-S has a dark matter halo that is fairly
round but a light distribution that is flattened (such as an edge-on
disk).  Alternatively, perhaps the two galaxies are interacting and
the relationship between light and mass is looser than for isolated
galaxies, with the presence of tidal tails for example.  In such a
case, our assumption of ellipsoidal mass distributions would be
incorrect, and future studies of this lens would need to consider a
wider range of mass distributions.

\lens\ is the second galaxy-mass gravitational lens system with an
image multiplicity larger than four. The first such system was the
six-image lens B1359+154 produced by a compact group of three
galaxies (Rusin et al.\ 2001). In both cases, the
high multiplicity is due to the presence of more than one lens
galaxy. It is interesting to note that both systems have a very
compact arrangement of galaxies, smaller than the typical massive
ellipticals that dominate lensing. In \lens\ the two galaxies have
velocity dispersions $\sigma \sim 120\,\kms$ and a projected
separation of just $0\farcs4$ (2~$h^{-1}$~kpc at $z_l = 0.76$),
while in B1359+154 the three galaxies have $\sigma \sim 140$--$160\,\kms$
and a maximum projected separation of $0\farcs7$ (4~$h^{-1}$~kpc at
$z_l \sim 1$; Rusin et al.\ 2001). Although these are just two out
of nearly 100 lenses currently known, they represent a much larger
fraction of the radio surveys that discovered them; B1359+154 is
one of 22 lenses in the Cosmic Lens All-Sky Survey (see Myers et
al.\ 1999; Browne et al.\ 2001), while \lens\ comes from a survey
that has found four lenses and one other candidate (Winn et al.\
2000, 2001, 2002a, 2002b). The incidence of high-multiplicity
lenses is still subject to Poisson uncertainties, but it appears to
be several percent or even higher. Lensing, especially radio lens
surveys, may therefore turn out to be an interesting way to
discover compact groupings of galaxies at redshifts out to
$z \sim 1$.

\acknowledgements It is a pleasure to thank Hans-Walter Rix, Paul
Schechter, and the other organizers of the 2002 Ringberg Workshop on
Gravitational Lensing, where this work commenced.  We thank Chris
Kochanek for interesting discussions, and the anonymous referee for
helpful comments on the manuscript.  C.R.K.\ thanks Pat Hall for
originally bringing \lens\ to his attention. C.R.K.\ is supported
by NASA through Hubble Fellowship grant HST-HF-01141.01-A from the
Space Telescope Science Institute, which is operated by the
Association of Universities for Research in Astronomy, Inc., under
NASA contract NAS5-26555. J.N.W.\ is supported by an NSF Astronomy \&
Astrophysics Postdoctoral Fellowship, under grant AST-0104347.

\clearpage

\appendix

\section{Handling Linear Parameters}

The surface mass density of an isothermal ellipsoid lens galaxy can
be written as follows, in units of the critical density for
lensing, and in a coordinate system aligned with the major axis of
the ellipse:
\begin{equation}
  \kappa(x,y) = \frac{\Sigma(x,y)}{\Sigma_{\rm crit}}
  = \frac{\sqrt{1+q^2}}{2\sqrt{2}\,q}\
    \frac{b}{\sqrt{x^2+y^2/q^2}}\ ,
\end{equation}
where $q = 1-e \le 1$ is the axis ratio of the ellipse, and $b$ is
a mass parameter. For a spherical galaxy, $b$ equals the Einstein
radius and is related to the velocity dispersion $\sigma$ of the
galaxy as
\begin{equation}
  b = 4\pi \left(\frac{\sigma}{c}\right)^2 \frac{D_{ls}}{D_{os}}\ ,
  \label{eq:bSIS}
\end{equation}
where $D_{ls}$ and $D_{os}$ are, respectively, angular diameter
distances from the lens and observer to the source. For a
non-spherical galaxy, this relation is modified by an
ellipticity-dependent factor of order unity (see Keeton et al.\
1997). Note that the surface mass density, and hence the lensing
potential and deflection, is linear in $b$. This leads to an
expression for the deflection of the form (see Kassiola \& Kovner
1993; Kormann et al.\ 1994; Keeton \& Kochanek 1998)
\begin{equation}
  \aa(\xx;b,\xx_g,e,\theta_e) = b\,\ah(\xx;\xx_g,e,\theta_e)\,,
\end{equation}
where $\ah$ is some function of position $\xx$ in the image plane
that depends on the parameters $\xx_g$, $e$, and $\theta_e$ (the
centroid, ellipticity, and orientation of the ellipsoid).

If we map an observed image $\xx_i$ to the source plane using the
two-galaxy-plus-shear model, we find the corresponding source
position
\begin{equation}
  \uu_i = \xx_i - b_1\,\ah_1(\xx_i;\xx_{g1},e_1,\theta_{e1})
    - b_2\,\ah_2(\xx_i;\xx_{g2},e_2,\theta_{e2})
    - \Gamma\cdot\xx_i\,,
\end{equation}
where the two components $\gamma_c$ and $\gamma_s$ of the shear
enter through the tensor
\begin{equation}
  \Gamma = \left[\begin{array}{rr}
    \gamma_c &  \gamma_s \\
    \gamma_s & -\gamma_c \\
  \end{array}\right] .
\end{equation}
We might imagine that instead of evaluating $\chi^2$ by
comparing the observed and predicted images in the image plane
(eq.~\ref{eq:chiimg}), we could simply work in the source plane
and use the scatter in the model source positions, defining
\begin{equation}
  \chi^2_{\src} = \sum_{i=1}^{N_{\rm images}}
    \frac{|\uu_i-\uu_{\mod}|^2}{\sigma_{i,x}^2}\ ,
    \label{eq:chisrc}
\end{equation}
where $\uu_{\mod}$ is the model source position. This
``source-plane $\chi^2$'' is often a good approximation to the
image-plane $\chi^2$ and can be valuable for modeling (see Kochanek
1991a). We do not use it as the basis of all modeling mainly
because it does not have the ability to check whether the model
predicts the correct number of images, which is essential in the
case of \lens.

The important point is that the parameters $b_1$, $b_2$,
$\gamma_c$, $\gamma_s$, and $\uu_{\mod}$ are all {\it linear\/}
parameters in $\chi^2_{\src}$. It is therefore possible to write
down a system of linear equations for the parameter values that
minimize $\chi^2_{\src}$. These equations have the form
\begin{equation}
  \MM\cdot\pp = \vv\,,
\end{equation}
where $\pp = (b_1,b_2,\gamma_c,\gamma_s,u_{\mod},v_{\mod})$ is the
parameter vector, and the matrix and right-hand side vector are:
\begin{equation}
  \MM = \sum_{i=1}^{N_{\rm images}}\frac{1}{\sigma_{i,x}^2}
  \left[\begin{array}{cccccc}
    \ah_{1x}^2 + \ah_{1y}^2 &
&&&&\\
    \ah_{1x}\ah_{2x} + \ah_{1y}\ah_{2y} &
    \ah_{2x}^2 + \ah_{2y}^2 &
&&&\\
    x_i\,\ah_{1x} - y_i\,\ah_{1y} &
    x_i\,\ah_{2x} - y_i\,\ah_{2y} &
    x_i^2 + y_i^2 &
&&\\
    y_i\,\ah_{1x} + x_i\,\ah_{1y} &
    y_i\,\ah_{2x} + x_i\,\ah_{2y} &
    0 &
    x_i^2 + y_i^2 &
&\\
    \ah_{1x} &
    \ah_{2x} &
    x_i &
    y_i &
    1 &
\\
    \ah_{1y} &
    \ah_{2y} &
    -y_i &
     x_i &
    0 &
    1
\\
  \end{array}\right]
\end{equation}
\begin{equation}
  \vv = \sum_{i} \frac{1}{\sigma_{i,x}^2} \left[\begin{array}{c}
    x_i\,\ah_{1x} + y_i\,\ah_{1y} \\
    x_i\,\ah_{2x} + y_i\,\ah_{2y} \\
    x_i^2 - y_i^2 \\
    2 x_i y_i \\
    x_i \\
    y_i \\
  \end{array}\right]
\end{equation}
where $(\ah_{1x},\ah_{1y})$ and $(\ah_{2x},\ah_{2y})$ are the $x$
and $y$ components of the (scaled) deflections for galaxy 1 and 2,
respectively, evaluated at the appropriate image position $\xx_i$;
and the sum is over the images. All the elements of $\MM$ can be
derived from the ones presented here, because $\MM$ is symmetric.

Using standard methods to solve this system of equations (e.g.,
Press et al.\ 1992), we can find the optimal values of the
parameters $\pp$ given any values of the other model parameters.
This technique can be used to aid the optimization as discussed in
\S3. In practice, we use only the values for $b_1$, $b_2$,
$\gamma_c$, and $\gamma_s$ recovered from this analysis, even
though solving the system of equations yields an estimate of the
source position as well.

\begin{deluxetable}{cccccrcrrrr}
\tablewidth{0pt}
\tablecaption{Sample Model Parameters}
\tablehead{
  \colhead{Type} &
  \colhead{$b$ (\arcsec)} &
  \colhead{$x_g$ (\arcsec)} &
  \colhead{$y_g$ (\arcsec)} &
  \colhead{$e$} &
  \colhead{$\theta_e$ ($^\circ$)} &
  \colhead{$\gamma$} &
  \colhead{$\theta_\gamma$ ($^\circ$)} &
  \colhead{$\chi^2_{\rm pos}$} &
  \colhead{$\chi^2_{\rm flux}$} &
  \colhead{$\chi^2_{\rm shape}$}
}
\startdata
%Model #2
assisted
& 0.15 & 0.50 & 0.18 & 0.64 & $ 11.5$ & 0.34 & $ 96.3$ &  0.09 & 13.06 & 2.58 \\
& 0.14 & 0.25 & 0.47 & 0.86 & $  4.6$ &      &         &       &       &      \\
%Model #7
assisted
& 0.16 & 0.49 & 0.19 & 0.54 & $ 15.4$ & 0.27 & $ 96.0$ &  0.04 & 14.07 & 0.58 \\
& 0.16 & 0.25 & 0.49 & 0.80 & $  5.2$ &      &         &       &       &      \\
%Model #10
assisted
& 0.20 & 0.45 & 0.19 & 0.09 & $108.2$ & 0.05 & $ 66.7$ &  0.03 & 16.38 & 5.59 \\
& 0.19 & 0.25 & 0.52 & 0.33 & $ 16.9$ &      &         &       &       &      \\
%Model #14
assisted
& 0.18 & 0.47 & 0.21 & 0.40 & $ 49.3$ & 0.11 & $113.5$ &  0.03 & 15.35 & 7.56 \\
& 0.17 & 0.26 & 0.53 & 0.66 & $ 21.9$ &      &         &       &       &      \\
%Model #20
assisted
& 0.18 & 0.47 & 0.20 & 0.29 & $ 36.3$ & 0.11 & $ 98.1$ &  0.01 & 15.18 & 2.06 \\
& 0.18 & 0.25 & 0.52 & 0.59 & $ 13.1$ &      &         &       &       &      \\
%Model #5
direct
& 0.17 & 0.48 & 0.19 & 0.40 & $  4.6$ & 0.18 & $ 85.6$ &  0.02 & 14.17 & 2.85 \\
& 0.18 & 0.23 & 0.48 & 0.62 & $178.5$ &      &         &       &       &      \\
%Model #11
direct
& 0.19 & 0.46 & 0.20 & 0.10 & $ 38.6$ & 0.06 & $ 83.1$ &  0.08 & 16.16 & 3.06 \\
& 0.19 & 0.24 & 0.51 & 0.41 & $ 12.4$ &      &         &       &       &      \\
%Model #17
direct
& 0.18 & 0.47 & 0.20 & 0.21 & $  9.6$ & 0.11 & $ 82.8$ &  0.02 & 15.46 & 1.22 \\
& 0.19 & 0.24 & 0.50 & 0.51 & $  2.9$ &      &         &       &       &      \\
%Model #21
direct
& 0.19 & 0.46 & 0.20 & 0.10 & $ 50.0$ & 0.06 & $ 84.5$ &  0.03 & 15.64 & 4.08 \\
& 0.19 & 0.25 & 0.51 & 0.44 & $ 13.3$ &      &         &       &       &      \\
%Model #22
direct
& 0.19 & 0.47 & 0.20 & 0.25 & $ 42.7$ & 0.09 & $102.1$ &  0.02 & 15.52 & 3.75 \\
& 0.18 & 0.25 & 0.52 & 0.56 & $ 18.8$ &      &         &       &       &      \\
\enddata
\tablecomments{
Parameters for five assisted and five direct models, randomly chosen
among models consistent with the 15~GHz image shapes (see \S4.4).
The mass parameter $b$ is defined in \refeq{bSIS} in the Appendix.
The positions $(x_g,y_g)$ of the galaxies are measured relative
to component D. The ellipticity $e$ and shear $\gamma$ are
dimensionless. The orientation angles $\theta_e$ and
$\theta_\gamma$ are given as position angles measured East of
North. The goodness of fit statistics $\chi^2_{\rm pos}$,
$\chi^2_{\rm flux}$, and $\chi^2_{\rm shape}$ refer to the image
positions, the flux ratios, and the image shapes, respectively.
}
\end{deluxetable}

\end{document}